\documentclass[
 reprint,
 amsmath,amssymb,
 aps,
]{revtex4-1}

\usepackage{graphicx}
\usepackage{dcolumn}
\usepackage{bm}
\usepackage{braket,amsmath,amssymb,bm}

\begin{document}

\preprint{APS/123-QED}

\title{Entanglement Generation by Communication using Phase-Squeezed Light with Photon Loss}

\author{Fumiaki Matsuoka}
\email{matsuoka@optnet.ist.hokudai.ac.jp}

\author{Akihisa Tomita}
 \email{tomita@ist.hokudai.ac.jp}

\author{Atsushi Okamoto}

\affiliation{
Graduate School of Information Science and Technology, Hokkaido University, Kita14-Nishi9, Kita-ku, Sapporo 060-0814, Japan
}

\begin{abstract}
To implement fault-tolerant quantum computation, entanglement generation with low error probability and high success probability is required. In a previous paper, we proposed the use of squeezed coherent light as a probe to generate entanglement between two atoms by communication, and showed that the error probability is reduced well below the threshold of fault-tolerant quantum computation [Phys. Rev. A. 88, 022313 (2013)]. In this paper, we investigate the effect of photon loss mainly due to finite coupling efficiency to the cavity. The error probability with photon loss is calculated using a beam-splitter model for homodyne measurement of probe light. We examine the optimum conditions of the amplitude of the probe light and the degree of squeezing to minimize the error probability. We show that the phase-squeezed probe light yields lower error probability than a coherent-light probe, even with photon losses. A fault-tolerant quantum computation algorithm can be implemented under 87\% transmittance by concatenating a seven-qubit error correction code for the phase flip error.
\begin{description}
\item[PACS numbers] 03.65.Ud, 03.67.Lx, 03.67.Hk, 42.50.Dv
\end{description}
\end{abstract}

\pacs{Valid PACS appear here}

\maketitle

\section{Introduction}
Quantum entanglement is essential for the implementation of quantum computation schemes \cite{01,02,03,04}. In 2001, one-way quantum computation \cite{02,05} was proposed to work out the complexities of quantum gate circuits. Although one-way quantum computation requires a large-scale cluster state before computation, it can be implemented via measurements on a quantum bit (qubit) and one-qubit unitary transformations according to the measurement outcomes. It has been reported from experiments that a cluster state (where $10^{5}$ atoms are considered to be entangled) can be generated by collisions of atoms in an optical lattice \cite{06}. However, quantum computation is difficult using this cluster state because the distance between the atoms is too small to select a single atom for one-qubit operation. Therefore, keeping sufficient distance between entangled atoms is important to implement one-way quantum computation.\par
Entanglement generation by communication (which uses an electromagnetic field, i.e., a quantum bus or qubus; see Sec. II for a more detailed description) has been proposed \cite{07,08,09,10} as an efficient entangler that can be used for creating cluster states \cite{11,12,13,14}. Although this technique can entangle distant atoms, the estimated error probability of entanglement generation is too large for any fault-tolerant quantum computation schemes. To reduce the error probability, we proposed the use of squeezed coherent light instead of coherent light as a qubus and showed that fault-tolerant one-way quantum computing \cite{15} can be implemented using phase-squeezed light \cite{16}. For this proposal, we assumed that the transmission loss for squeezed light was negligible because the distances between quantum memories are small. However, photon loss may also result from coupling to the cavities and the measurement of the squeezed coherent light. The travelling wave light is expected to experience a loss at the interface to a typical Fabry--P$\acute{{\text e}}$rot cavity \cite{17}. Moreover, the efficiency of the homodyne measurement is less than unity, owing to the finite quantum efficiencies of the photodetectors as well as mode mismatch between a signal and a local oscillator \cite{18}. The phase-squeezed light may be collapsed by these losses, and tends to become coherent light.\par
In this report, we estimate the error probability of entanglement generation using squeezed coherent light under photon loss. The effect of photon loss on entanglement generation with squeezed light in transmission has been studied for qubuses in the context of quantum repeaters \cite{19}. The use of squeezed light and the displacement operation has been predicted to improve the fidelity and the success probability of dispersive interaction for coherent light from 0.89 to 0.77 and from 40\% to 36\%, respectively, for a node interval of 10 km \cite{19}. Meanwhile, entanglement generation for quantum computers requires a different characterization than that for quantum repeaters. One cannot directly apply the results in \cite{19} to entanglement generation for quantum computation. Fidelity of entanglement between two atoms is important for quantum repeaters since only the availability of entanglement purification is a concern \cite{20}. On the other hand, error probability in the gates is important for implementing fault-tolerant quantum computation. When the success probability of the quantum gate is provided, the error probability of the quantum gate should be below the threshold values for fault-tolerance \cite{21,22,23,24}. For example, an error probability no greater than $4 \times 10^{-4}$ is required for a success probability of 50\% in the Fujii and Tokunaga (FT) method \cite{15}, which is based on a topological code. In another example, an error probability no greater than $1 \times 10^{-2}$ is required for a success probability of 100\% in Knillfs method \cite{25}, which is based on error-correcting teleportation. Note that fidelity refers to the distance between the ideal Bell state and the generated entangled state \cite{07}, whereas error probability refers to the probability of an unexpected error in entanglement generation. In a qubus, this error originates from overlaps between non-orthogonal probe light states \cite{08}. In addition, most fault-tolerant quantum computing architectures \cite{25,26,27,28} require a success probability of at least 1/2, while quantum repeaters allow for a lower success probability. In fact, Praxmeyer and van Loock \cite{19} employed homodyne measurements and window functions for quantum state discrimination to obtain the final state with a high fidelity to the ideal Bell state, which reduced the success probability to less than 1/2. In the following analysis for fault-tolerant quantum computation, we calculate the error probability with only homodyne measurements to obtain a success probability equal to 1/2.\par
In addition to the collapse of the phase-squeezed light, we should consider phase flip errors for atoms induced by the photon loss. In the beam splitter model, which is known to be a simple approach for photon loss, transmitted and reflected (lost) photons provide gwhich-pathh information. Since atoms and probe light are entangled, phase flip error is induced by tracing over the lost photon modes \cite{10}. The total error probability, which thus consists of the error originating from the homodyne measurement and the phase flip error induced by photon loss, may be an obstacle to implementing fault-tolerant quantum computing. To address this problem, we introduce three-qubit and seven-qubit quantum error correction codes (QECC) for the phase flips.\par
The remainder of this paper is structured as follows. We briefly review entanglement generation by communication using squeezed coherent light in Sec. II. Then, we show a homodyne measurement scheme for three-state discrimination and calculate the error probability with photon loss of the squeezed coherent light in Sec. III. In Sec. IV, we introduce the phase error correcting code for phase flip error and examine the feasibility of the implementation of fault-tolerant quantum computation. Finally, we conclude in Sec. V.

\section{Entanglement Generation by Phase-Squeezed Light and Homodyne Measurement}
In this section, we briefly review entanglement generation between two atoms by squeezed light \cite{16}. Figure 1(a) shows the process of entanglement generation with a qubus. The process consists of sequential atom-probe interactions and measurements. Two atoms have lower energy states $\ket{0}$ and $\ket{1}$ and an excited state $\ket{e}$, where only the transition $\ket{1} \longleftrightarrow \ket{e}$ is allowed \cite{29,30,31}, as shown in Fig. 1(b). We assume dispersive atom-photon interaction \cite{08,09,31} by a large detuning $\Delta$ and a not-too-large mean photon number $\bar{n}$ in the cavity mode. Then, the atom-photon interaction provides a unitary operator that conditionally rotates the phase of the photon as:
\begin{eqnarray}
\hat{U}=\hat{I}\ket{0}\bra{0}+\hat{R}(\theta)\ket{1}\bra{1},
\end{eqnarray}
where $\theta=\chi t$, $\chi = g^2 / \Delta $, $g$ is a coupling constant, and $\hat{R} (\theta)=e^{i\theta \hat{n}}$. The probe light is the squeezed coherent state $\ket{\xi, \alpha}$, where $\xi =r_{0} e^{i \varphi}$ is the squeezing parameter, described by amplitude $r_{0}$ and phase $\varphi$. In addition, we fix the mean photon number $\bar{n}$ between the coherent state $\ket{\alpha}$ and the squeezed coherent state $\ket{\xi, \beta}$ for a fair comparison by adjusting the amplitude to $\beta$ from the original value of $\alpha$, where $\beta =\left| \sqrt{|\alpha|^{2}-\text{sinh}^{2}(r)} \right|$. We use the phase $\varphi=\pi$ since it is the optimal squeezing direction for quadrature fluctuations in a homodyne measurement \cite{16}. For $\varphi=\pi$, phase fluctuations of quadrature are squeezed; thus, this is called phase-squeezed light $\ket{r, \beta}$, where $r=-r_{0}$.\par
The interaction between the light and atom A creates an entanglement between the atomic states and the light states as $\frac{1}{\sqrt{2}} (\ket{0}\ket{r, \beta}+\ket{1}\ket{re^{i2\theta}, \beta e^{i\theta}})$ from the initial product state $\frac{1}{\sqrt{2}}(\ket{0}+\ket{1})\otimes \ket{r, \beta}$ by the unitary operation of Eq. (1). Then, the interaction between the light and atom B yields the final state:
\begin{eqnarray}
\ket{\psi_{2}}&=& \frac{1}{2}\ket{0}\ket{0}\ket{r, \beta}+\frac{1}{\sqrt{2}}\left( \frac{\ket{0}\ket{1}+\ket{1}\ket{0}}{\sqrt{2}}\right) \ket{re^{i2\theta}, \beta e^{i\theta}} \nonumber \\
&& +\frac{1}{2}\ket{1}\ket{1}\ket{re^{i4\theta}, \beta e^{i2\theta}}.
\end{eqnarray} 
After that, the probe light is measured. Although we have previously estimated the error probability with minimum error discrimination \cite{16}, this may be difficult to implement for three squeezed coherent states. For a more feasible situation, we instead consider a homodyne measurement of the probe light. The entangled state of the atoms $\ket{0}\ket{1}+\ket{1}\ket{0}$ is formed by post-selecting the rotated state $\ket{re^{i2\theta}, \beta e^{i\theta}}$. A homodyne measurement projects the quantum state onto a projection axis. Then, as we want to discriminate the $\ket{re^{i2\theta}, \beta e^{i\theta}}$ state from the other two states in Eq. (2), the projection axis should be taken to be the ($p+\theta$)-axis for the optimal phase measurement. In this situation, the probability density distributions, as shown in Fig. 2(b), can be obtained corresponding to the measurement outcome of the probe light \cite{16}. The error in entanglement generation in homodyne measurement occurs as a result of overlaps in the probability density distributions representing the measurement outcomes.
\begin{figure*} [htbp]
\includegraphics[width=10cm]{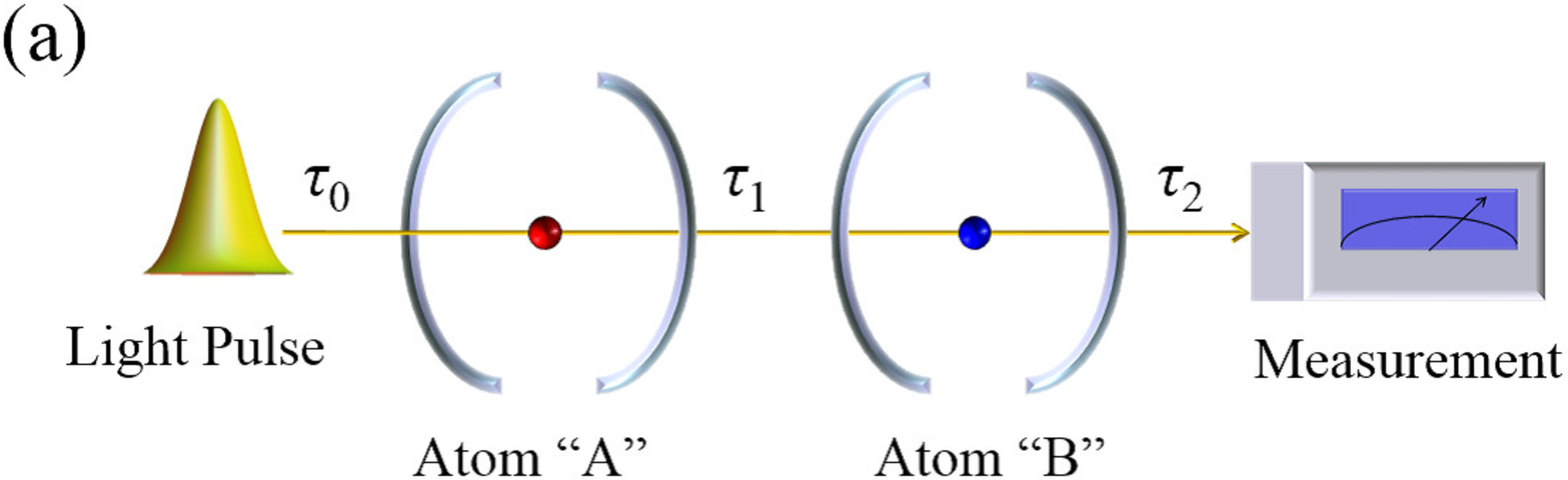}  \includegraphics[width=5cm]{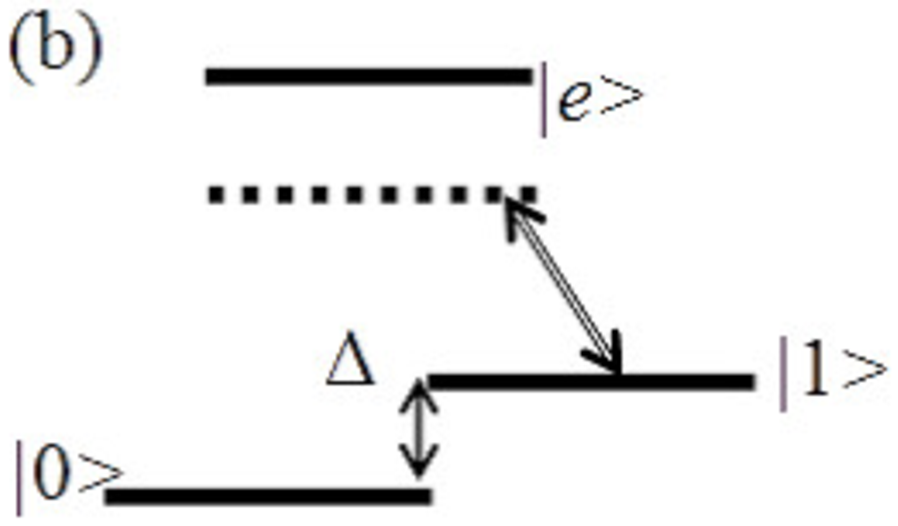}
\caption{\label{fig:wide} (Color online) (a) The process of entanglement generation between quantum memories by communication. Light interacts with two atoms, and the phase rotation caused by the interaction is measured after the interaction. (b) The energy level scheme of a $\Lambda$-configured three-level atom. The phase of the light rotates only when the atom is in state $\ket{1}$.}
\end{figure*}
\begin{figure} [htbp]
\includegraphics[width=6cm]{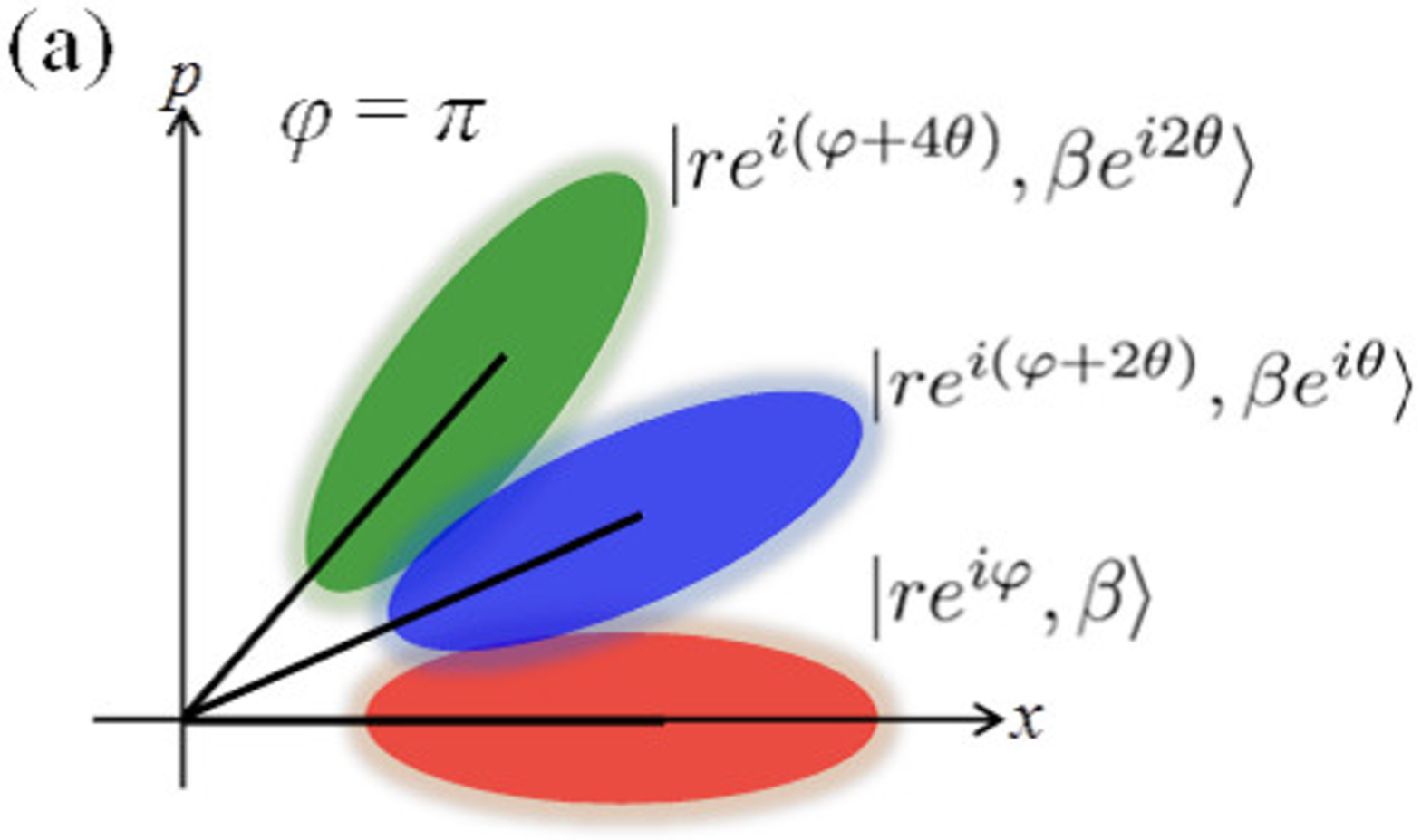}  \\
\includegraphics[width=7cm]{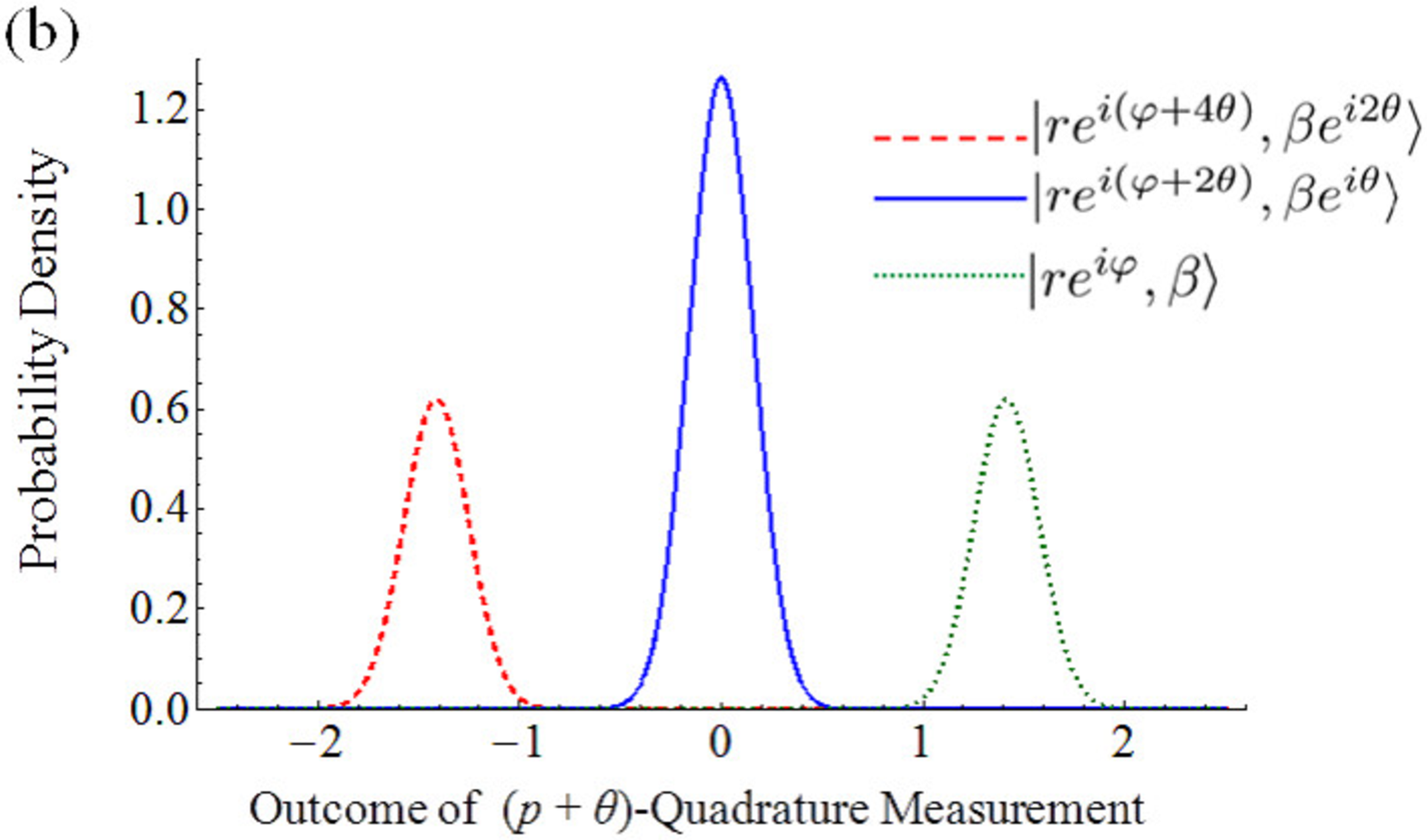}
\caption{\label{fig2} (Color online) (a) A schematic representation of phase-squeezed light in a phase space after interacting with atoms A and B (at time $\tau_{2}$ in Fig. 1(a)), described by ellipses. The atomic entanglement state $\ket{0}\ket{1}+\ket{1}\ket{0}$ is formed by post-selecting the rotated coherent state $\ket{re^{i2\theta}, \beta e^{i\theta}}$. (b) Probability density distributions for the outcomes of quadrature measurements of the probe beam corresponding to the quantum states in Fig. 2(a) projected on the ($p+\theta$)-axis (where $p$ is the imaginary part of $\alpha$), which is the optimal projection angle to obtain the lowest error probability in the homodyne measurement.}
\end{figure}

\section{Error Probability in Homodyne Measurement with Loss}
In this section, we examine the effects of loss on the error probability of entanglement generation. As mentioned earlier, we calculate the overlaps of the probability density distributions for homodyne detection of phase-squeezed light \cite{32}.\par
Photon loss can be described by a beam splitter model \cite{18} characterized by transmittance $\eta$ ($0<\eta<1$), as shown in Fig. 3. The loss is described by $1-\eta$. In our scheme, although photon loss may be induced in transmission, cavity coupling, and measurement of the probe light, such losses can be represented by a single beam splitter whose reflection corresponds to the total photon loss because the beam splitter operation and the phase rotation operation commute \cite{33}. In practical terms, since the phase shift angle may be decreased by photon loss, either a displacement operation before the second cavity or an adjustment of the atom-light interaction time in the second cavity is required. In this report, we assume the phase shifts of the two cavities are equivalent by adjustment of the interaction time.\par
The probability density distributions when phase-squeezed states are projected onto the $x_{\lambda}$-axis (where $\lambda$ is the projective angle) are Gaussian with mean $\left< x_{\lambda} \right>$ and variance $\Delta x_{\lambda}^{2}$ \cite{34}. Note that the expectation value of $\hat{x}_{\lambda}$ with photon losses \cite{32} is  $\sqrt{\eta}$ times Eq. (21) of Ref. \cite{16}, and so the variance of $\hat{x}_{\lambda}$ is \cite{32}:
\begin{eqnarray}
\Delta x_{\lambda}^{2}&\equiv & \left< \hat{x}_{\lambda}^{2} \right> -\left< \hat{x}_{\lambda}\right>^{2} \nonumber \\
&=& \frac{\eta}{2} \left\{ \text{exp}(2r) \text{sin}^{2}\left( \lambda -\frac{\varphi}{2}+\theta \right) \right. \nonumber \\
&& \left. +\text{exp}(-2r)\text{cos}^{2} \left( \lambda -\frac{\varphi}{2}+\theta \right) \right\}+\frac{1-\eta}{2}.
\end{eqnarray} 
Here, $\left< \hat{x}_{\lambda}^{2} \right>$ is the expectation value of $x_{\lambda}^{2}$. The photon loss decreases the amplitude parameter by $\sqrt{\eta}$ and reduces the squeezing effect, as shown in the variance given by Eq. (3). In addition, we introduce the phase shift $\theta$ from the projective angle  $\lambda$ to Eqs. (19) and (21) of Ref. \cite{16} (the probability density distributions and the expectation value of a quadrature operator, respectively), and Eq. (3) to calculate the overlaps.\par
Using Eq. (3), and Eqs. (19) and (20) of \cite{16}, and assuming the direction of the projection axis is $\lambda=(\pi+\theta)/2$ (projection onto the $p+\theta$-axis in Fig. 2(a), which corresponds to phase measurement), we obtain the probability density distributions of the measurement outcomes on the superposition state described by Eq. (2), as shown in Fig. 4. The overlaps of the probability density distributions are increased by the photon loss compared with those calculated for a lossless case as depicted in Fig. 2(b). \par
The error probability is obtained rigorously by subtracting the overlap between the probability density distributions of $\ket{r, \beta}$ and $\ket{re^{i4\theta}, \beta e^{i2\theta}}$ from the sums of the overlaps between  $\ket{r, \beta}$ and $\ket{re^{i2\theta}, \beta e^{i\theta}}$ and the overlaps between $\ket{re^{i2\theta}, \beta e^{i\theta}}$ and $\ket{re^{i4\theta}, \beta e^{i2\theta}}$. These overlaps can be calculated using Eqs. (24)-(27) of Ref. \cite{16}, where the expectation value $\left< x_{\lambda} \right>$, variance $\Delta x^{2}_\lambda$, and the coherent amplitude on the integral interval $\alpha^{''}$can be replaced by  $\sqrt{\eta}\left< x_{\lambda} \right>$, Eq. (3), and $\sqrt{\eta}\alpha^{''}$, respectively. Here, we examine the effect of photon loss on the error probability. Figure 5(a) plots the error probability as a function of the transmittance $\eta$ for various values of the squeezing amplitude $r$. In the lossless case ($\eta=1$) \cite{16}, for typical values of $\bar{n}=10^{4}$ and $\theta=0.01$ \cite{07,35,36}, we obtain the error probability of entanglement generation with the homodyne measurement as $P_{\text{E}}=0.23$ with squeezing amplitude $r=0$ (for coherent states) and with the experimentally reported maximum value of the squeezing parameter $r=1.5$ \cite{37}. As photon loss increases (i.e., for smaller transmittance $\eta$), the error probabilities increase and converge to 0.5 even for large squeezing amplitudes. This is because the overlaps between the three phase-squeezed states shown in Fig. 4 depend not only on the coherent amplitude, but also on the variance $\Delta x_{\lambda}^{2}$ given by Eq. (3). The squeezed variance is collapsed to the variance of coherent light $\Delta x_{\lambda}^{2}=1/2$, owing to the invasion of the vacuum mode by the photon loss. In fact, the variance  $\Delta x_{\lambda}^{2}$ for the phase (i.e., $p$-quadrature; $\Delta x_{\lambda+\pi/2}^{2}$) for phase-squeezed states increases as $\eta$ decreases, and reaches the value for the coherent state at $\eta=0$, as shown in Fig. 5 (b). Nevertheless, the phase-squeezed state provides a better error probability than the coherent state for phase measurement provided that $\eta>0$.
\begin{figure} [htbp]
\includegraphics[width=7cm]{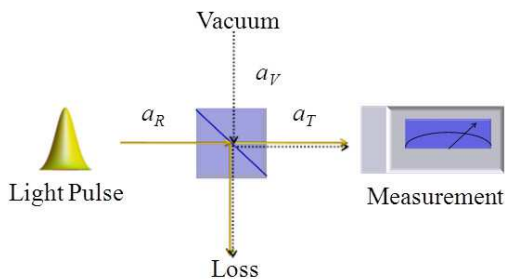}  
\caption{\label{fig3} (Color online) A schematic of the beam splitter model. The loss is represented by the reflection at the beam splitter with a transmittance $\eta$. When the probe light with the amplitude operator $a_{\text{T}}$ enters the beam splitter, the amplitude operators of the transmitted and reflected light become $\sqrt{\eta}a_{\text{T}}$ and $\sqrt{1-\eta}a_{\text{T}}$, respectively. On the other hand, when vacuum with an amplitude operator $a_{\text{V}}$ is placed at the other port of the beam splitter, the transmitted and reflected light become $\sqrt{1-\eta}a_{\text{V}}$ and $\sqrt{\eta}a_{\text{V}}$, respectively. Therefore, the transmitted light and vacuum are coupled to $a_{\text{R}}=\sqrt{\eta}a_{\text{T}}+\sqrt{1-\eta}a_{\text{V}}$ by the beam splitter and the light has a perfect measurement.}
\end{figure}
\begin{figure} [htbp]
\includegraphics[width=7cm]{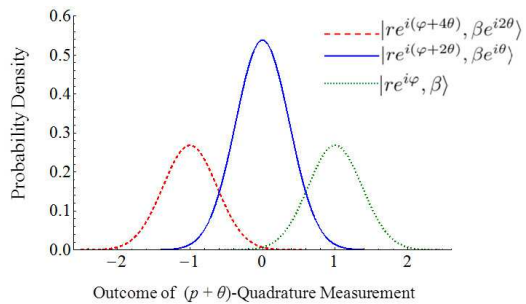}  
\caption{\label{fig4} (Color online) Probability density distributions of the superposition of the phase-squeezed state with losses. The overlaps between these distributions are increased by the loss effect compared with the lossless case, such as that shown in Fig. 2.}
\end{figure}
\begin{figure} [htbp]
\includegraphics[width=7cm]{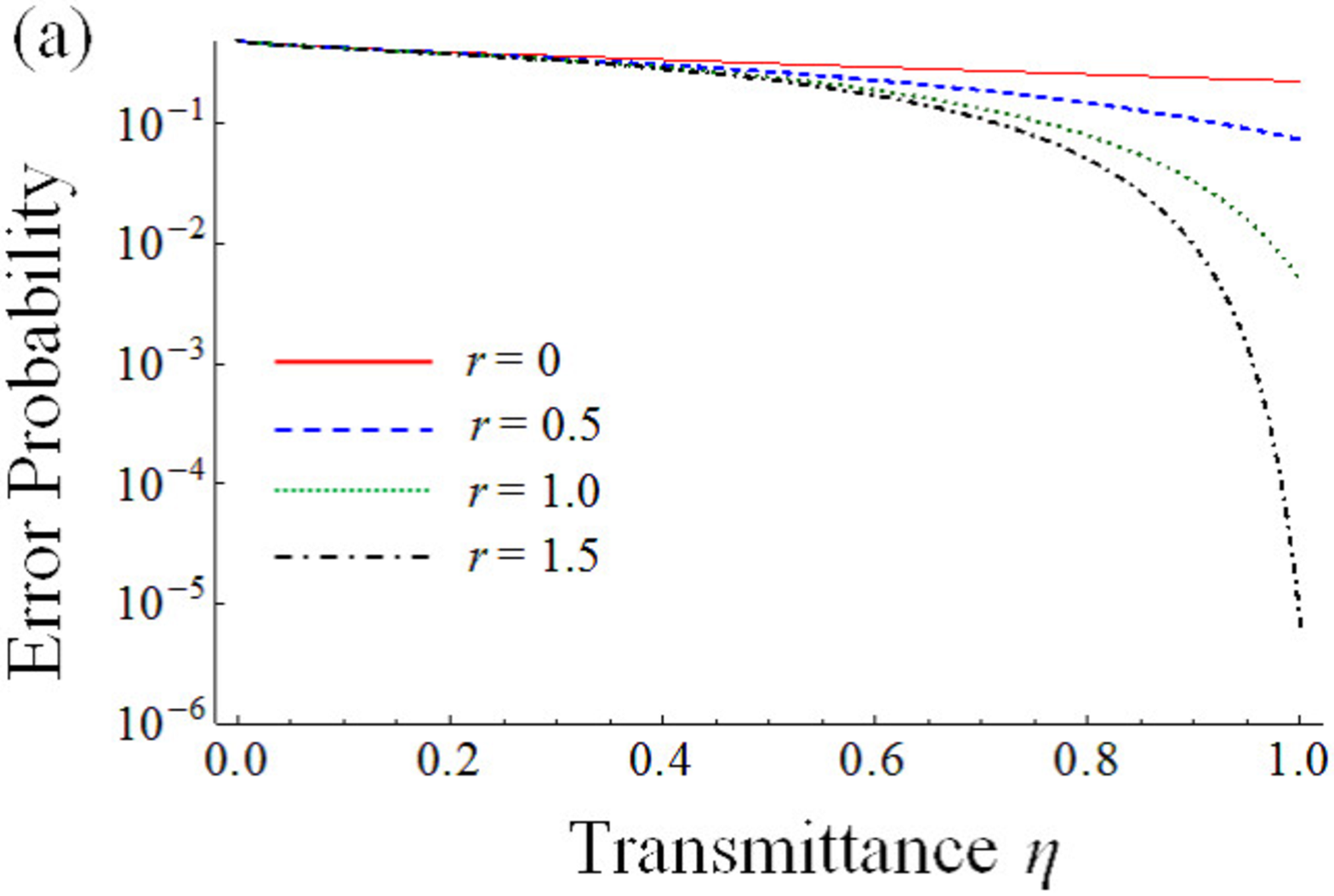}  \\
\includegraphics[width=7cm]{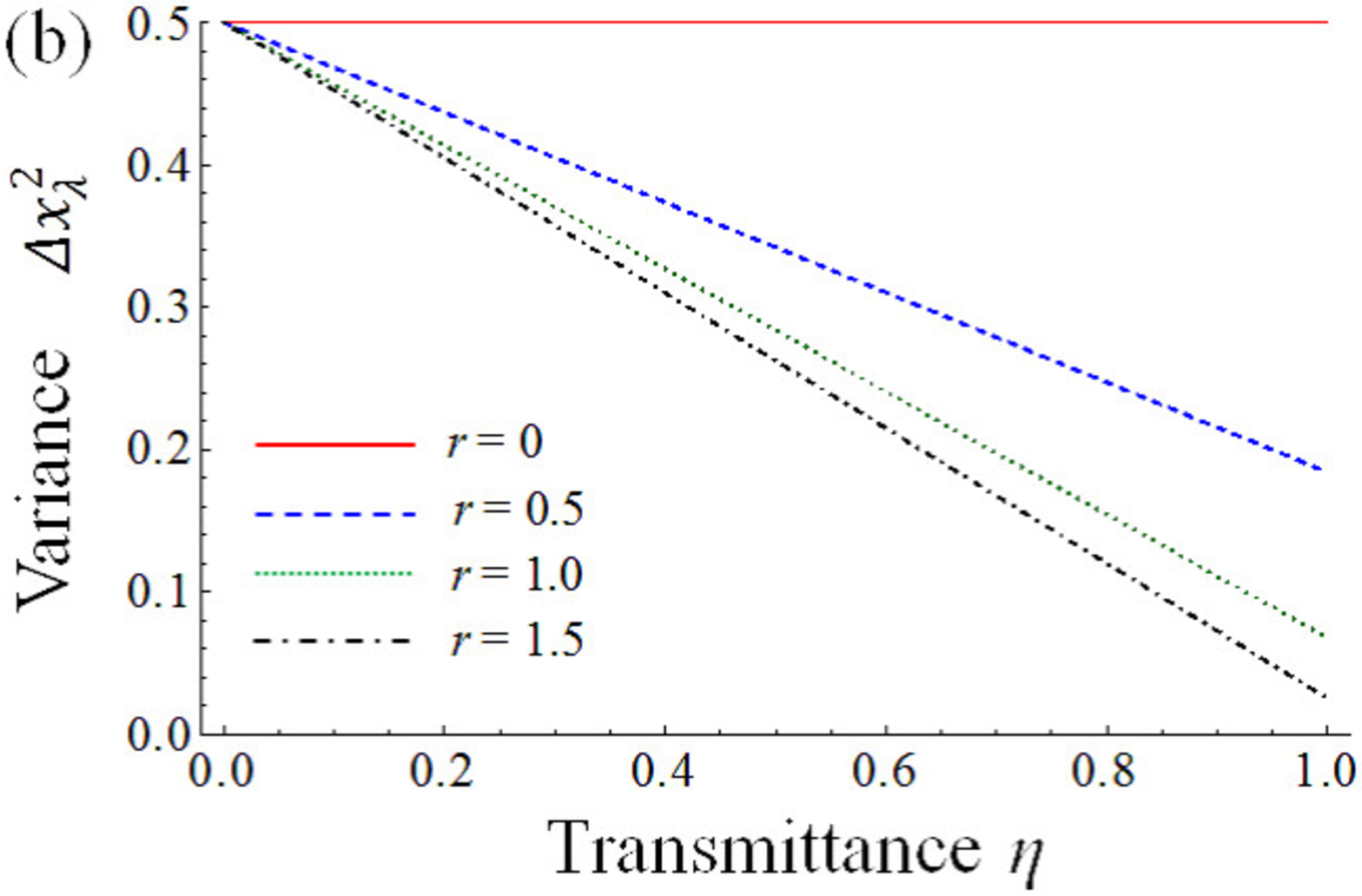}
\caption{\label{fig5} (Color online) (a) The error probability of entanglement generation and (b) the variance  $\Delta x_{\lambda}^{2}$ as a function of $\eta$ with squeezing amplitude $r=0$ (solid line, coherent state), $r=0.5$ (dashed line), $r=1.0$ (dotted line), and $r=1.5$ (dot-dashed line). In phase-squeezed light, the loss effect is significant in comparison to that of coherent light (see main text for details).}
\end{figure}

\section{Phase filp error induced by photon losses and error correcting codes}
\subsection{Error correction for phase flip error} 
We here examine the impact of photon loss on the implementation of fault-tolerant quantum computation in terms of the error probability. To this end, since atoms and probe light are entangled, we must consider the fact that photon loss induces phase flip error \cite{10} in addition to the measurement-induced error described in Sec. III. To confirm the effect of photon losses for each atom, we add two beam splitters, as shown in Fig. 6, since the effects of photon losses for atoms A and B are different. In this model, since the atoms and probe light are entangled at the beam splitters, the transmitted and reflected (lost) photons provide ``which-path'' information of the atoms. Since the reflected photon modes have the information of the atoms, the information of atoms is also lost.  Therefore, the phase flip errors of the atoms are induced by photon losses \cite{10}; thus, the final state becomes a mixed state. The phase flip error rate is determined by the inner product between the non-phase-shifted state $\ket{r, \beta}$ and the phase-shifted state $\ket{r e^{i2\theta}, \beta e^{i\theta}}$ \cite{10}. When photon losses are induced, such an inner product can be written as:
\begin{eqnarray}
\gamma(\eta) &=& \frac{\pi}{8}\int_{-\infty}^{\infty} \int_{-\infty}^{\infty} W(x,p) \nonumber \\
&& W(x\text{cos}\theta -p\text{sin}\theta, x\text{sin}\theta +p\text{cos} \theta)dxdp,  
\end{eqnarray} 
where $W(x,p)$ and $W(x\text{cos}\theta -p\text{sin}\theta, x\text{sin}\theta +p\text{cos} \theta)$ are Wigner functions of the non-phase-shifted and phase-shifted phase-squeezed states, respectively \cite{38}. The Wigner function with photon loss is given by:
\begin{eqnarray}
W(x,p) &=& N \text{exp}\left[ -\frac{2e^{-2 r}(x-\sqrt{\eta} x_{i})^{2}}{\eta + (1-\eta )e^{-2r}} \right. \nonumber \\
&& \left. -\frac{2(p-\sqrt{\eta} p_{i})^{2}}{\eta e^{-2r}+ (1-\eta )} \right],
\end{eqnarray} 
where  $x_{i}$ and $p_{i}$ represent the center of the phase-squeezed state ($x_{i}=\beta$ and $p_{i}=0$ for $W(x,p)$ and $W(x\text{cos}\theta -p\text{sin}\theta, x\text{sin}\theta +p\text{cos} \theta)$), and $N$ is the normalization constant:
\begin{eqnarray}
N &=& \frac{\pi}{2}\frac{1}{\sqrt{2}e^{-r}} \nonumber \\
&& \sqrt{(e^{-4r}+1)-(e^{-2r}-1)^{2}\{ (1-\eta)^{2}+\eta^{2} \}}.
\end{eqnarray} 
Note that since the transmission distance and the number of the cavity input-output process are different for atoms A and B, as shown in Fig. 6, photon loss effects for atoms A and B are also different; thus, we should consider the phase flip error probabilities in atoms A and B separately. For atom A, phase flip error is induced by photon losses from the first cavity to the detectors in the homodyne measurement. We assume the transmittance for atom A is $\eta_{\text{A}}=\eta_{1}\eta_{2}$, where $\eta_{\text{1}}$ and $\eta_{\text{2}}$ are transmittances from atom A to atom B, and from atom B to the detectors, respectively. In contrast, for atom B, phase flip error is induced by photon losses from the second cavity to the detectors. We assume that the transmittance for atom B is $\eta_{B}=(1-\eta_{1})\eta_{2}$. In the present analysis, we assume the same transmittances for $\eta_{\text{1}}$ and $\eta_{\text{2}}$, i.e., the total transmittance is given by $\eta=\eta_{\text{1}}\eta_{\text{2}}=\eta_{\text{1}}^{2}=\eta_{\text{2}}^{2}$. Using these transmittance values and Eqs. (4)-(6), the no-phase flip error probability $P_{\text{S}}$ and the phase flip error probability  $P_{\text{F}}$ for atoms A and B can be written as \cite{10}:
\begin{eqnarray}
P_{\text{S}}^{\text{A}}=\frac{1+\gamma(\eta_{A})}{2}, P_{\text{F}}^{\text{A}}=1-P_{\text{S}}^{\text{A}}, \\
P_{\text{S}}^{\text{B}}=\frac{1+\gamma(\eta_{B})}{2}, P_{\text{F}}^{\text{A}}=1-P_{\text{S}}^{\text{B}}.
\end{eqnarray} 
From these phase flip errors, the final state $\ket{\psi_{2}}$ can be rewritten as a mixed state \cite{10}:
\begin{eqnarray}
(P_{\text{S}}^{\text{A}}P_{\text{S}}^{\text{B}}+P_{\text{F}}^{\text{A}}P_{\text{F}}^{\text{B}})\ket{\psi_{2}}\bra{\psi_{2}}+P_{\text{F}}^{\text{A}}P_{\text{S}}^{\text{B}}\ket{\psi_{2}^{\text{flipA}}}\bra{\psi_{2}^{\text{flipA}}} \nonumber \\ 
+P_{\text{S}}^{\text{A}}P_{\text{F}}^{\text{B}}\ket{\psi_{2}^{\text{flipB}}}\bra{\psi_{2}^{\text{flipB}}}, 
\end{eqnarray} 
where states $\ket{\psi_{2}^{\text{flipA}}}$ and $\ket{\psi_{2}^{\text{flipB}}}$ are the phase-flipped states of $\ket{\psi_{2}}$ for atoms A and B, respectively. These states can be written respectively as:
\begin{eqnarray}
\ket{\psi_{2}^{\text{flipA}}}&=& \frac{1}{2}\ket{0}\ket{0}\ket{r, \beta} \nonumber \\
&& +\frac{1}{\sqrt{2}}\left( \frac{\ket{0}\ket{1}-\ket{1}\ket{0}}{\sqrt{2}}\right) \ket{re^{i2\theta}, \beta e^{i\theta}} \nonumber \\
&& -\frac{1}{2}\ket{1}\ket{1}\ket{re^{i4\theta}, \beta e^{i2\theta}},
\end{eqnarray}
and:
\begin{eqnarray}
\ket{\psi_{2}^{\text{flipB}}}&=& \frac{1}{2}\ket{0}\ket{0}\ket{r, \beta} \nonumber \\
&& +\frac{1}{\sqrt{2}}\left( \frac{-\ket{0}\ket{1}+\ket{1}\ket{0}}{\sqrt{2}}\right) \ket{re^{i2\theta}, \beta e^{i\theta}} \nonumber \\
&& -\frac{1}{2}\ket{1}\ket{1}\ket{re^{i4\theta}, \beta e^{i2\theta}}.
\end{eqnarray} 
By combining the error probability of the state discrimination, the total error probability is obtained as:
\begin{eqnarray}
E_{\text{tot}} &=& (P_{\text{S}}^{A}P_{\text{S}}^{B}+P_{\text{F}}^{A}P_{\text{F}}^{B})P_{\text{E}}+(P_{\text{F}}^{A}P_{\text{S}}^{B} \nonumber \\
&& +P_{\text{S}}^{A}P_{\text{F}}^{B})(1-P_{\text{E}}).
\end{eqnarray} 
\par
\begin{figure*} [htbp]
\includegraphics[width=12cm]{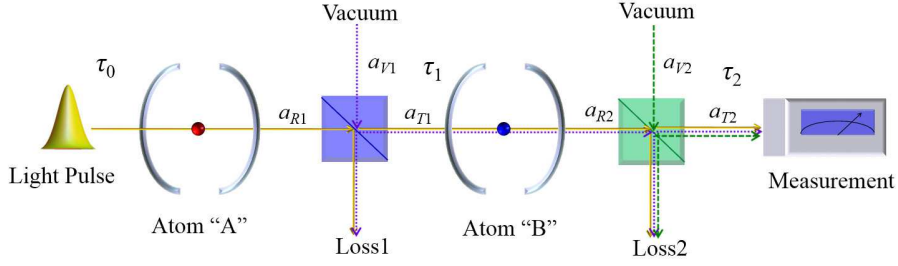}  
\caption{\label{fig:wide} (Color online) Schematic representation of photon losses of probe light for atoms A and B. Photon losses occur in the transmission of the probe light, the cavity input?output, and the measurement.}
\end{figure*}
Here, we examine the feasibility of fault-tolerant quantum computation schemes.  Knillfs method \cite{25} and FT method \cite{15} can be practically implemented with a realistic amount of resources when the error probability is below $1\times 10^{-2}$ and $4\times 10^{-4}$, respectively. Since the Knillfs method can tolerate a higher error probability than the FT method,  we examine the feasibility of Knillfs method. To examine this feasibility, we calculate the error probability $E_{\text{tot}}$ for $\theta=0.01$ as a function of the coherent amplitude $\alpha$ with the amplitudes of the following values: squeezing parameter $r=0$ and transmittance $\eta=0.9$ (solid line),   $r=0$ and $\eta=0.99$ (dotted line), $r=0.6$ and $\eta=0.9$ (dashed line), and  $r=0.8$ and $\eta=0.99$ (dot-dashed line), as shown in Fig. 7. Note that, since the phase flip error probability increases as the inner product Eq. (4) decreases, the total error probability can be minimized by optimization of the coherent amplitude  $\alpha$ and/or the phase shift angle $\theta$, and the squeezing parameter $r$. Even for small photon loss, the error probability  $E_{\text{tot}}\leq 1\times 10^{-2}$ is hard to realize, and the implementation of Knillfs method is difficult even if optimized phase-squeezed light is used.\par
To address this, we introduce the concatenation of a few-qubit QECC \cite{24} to correct the phase flip error. Such error corrections have been implemented in solid-state qubits, such as diamond spin systems \cite{39}. We estimate the probabilities of successful error correction on the logical errors. That is, the phase flip and no-phase flip error probabilities using a three-qubit QECC corresponding to atoms A and B are respectively:
\begin{eqnarray}
P_{\text{F3cor}}^{\text{A}}&=&(P_{\text{S}}^{\text{A}})^{3}+3(P_{\text{S}}^{\text{A}})^{2}P_{\text{F}}^{\text{A}}, \nonumber \\ P_{\text{S3cor}}^{\text{A}}&=&1-P_{\text{F3cor}}^{\text{A}}
\end{eqnarray} 
and:
\begin{eqnarray}
P_{\text{F3cor}}^{\text{B}}&=&(P_{\text{S}}^{\text{B}})^{3}+3(P_{\text{S}}^{\text{B}})^{2}P_{\text{F}}^{\text{B}}, \nonumber \\ P_{\text{S3cor}}^{\text{B}}&=&1-P_{\text{F3cor}}^{\text{B}}.
\end{eqnarray} 
Moreover, to obtain higher immutability to the photon loss, we introduce a seven-qubit QECC. Similar to the three-qubit QECC, the phase flip and no-phase flip error probabilities using the seven-qubit code for atoms A and B are respectively:
\begin{eqnarray}
P_{\text{F7cor}}^{\text{A}}&=&(P_{\text{S}}^{\text{A}})^{7}+7(P_{\text{S}}^{\text{A}})^{6}P_{\text{F}}^{\text{A}}+21(P_{\text{S}}^{\text{A}})^{5}(P_{\text{F}}^{\text{A}})^{2} \nonumber \\
&& +35(P_{\text{S}}^{\text{A}})^{4}(P_{\text{F}}^{\text{A}})^{3}, \nonumber \\ P_{\text{S7cor}}^{\text{A}}&=&1-P_{\text{F7cor}}^{\text{A}}
\end{eqnarray} 
and:
\begin{eqnarray}
P_{\text{F7cor}}^{\text{B}}&=&(P_{\text{S}}^{\text{B}})^{7}+7(P_{\text{S}}^{\text{B}})^{6}P_{\text{F}}^{\text{B}}+21(P_{\text{S}}^{\text{B}})^{5}(P_{\text{F}}^{\text{B}})^{2} \nonumber \\
&& +35(P_{\text{S}}^{\text{B}})^{4}(P_{\text{F}}^{\text{B}})^{3}, \nonumber \\ P_{\text{S7cor}}^{\text{B}}&=&1-P_{\text{F7cor}}^{\text{B}}.
\end{eqnarray} 
\begin{figure} [htbp]
\includegraphics[width=7cm]{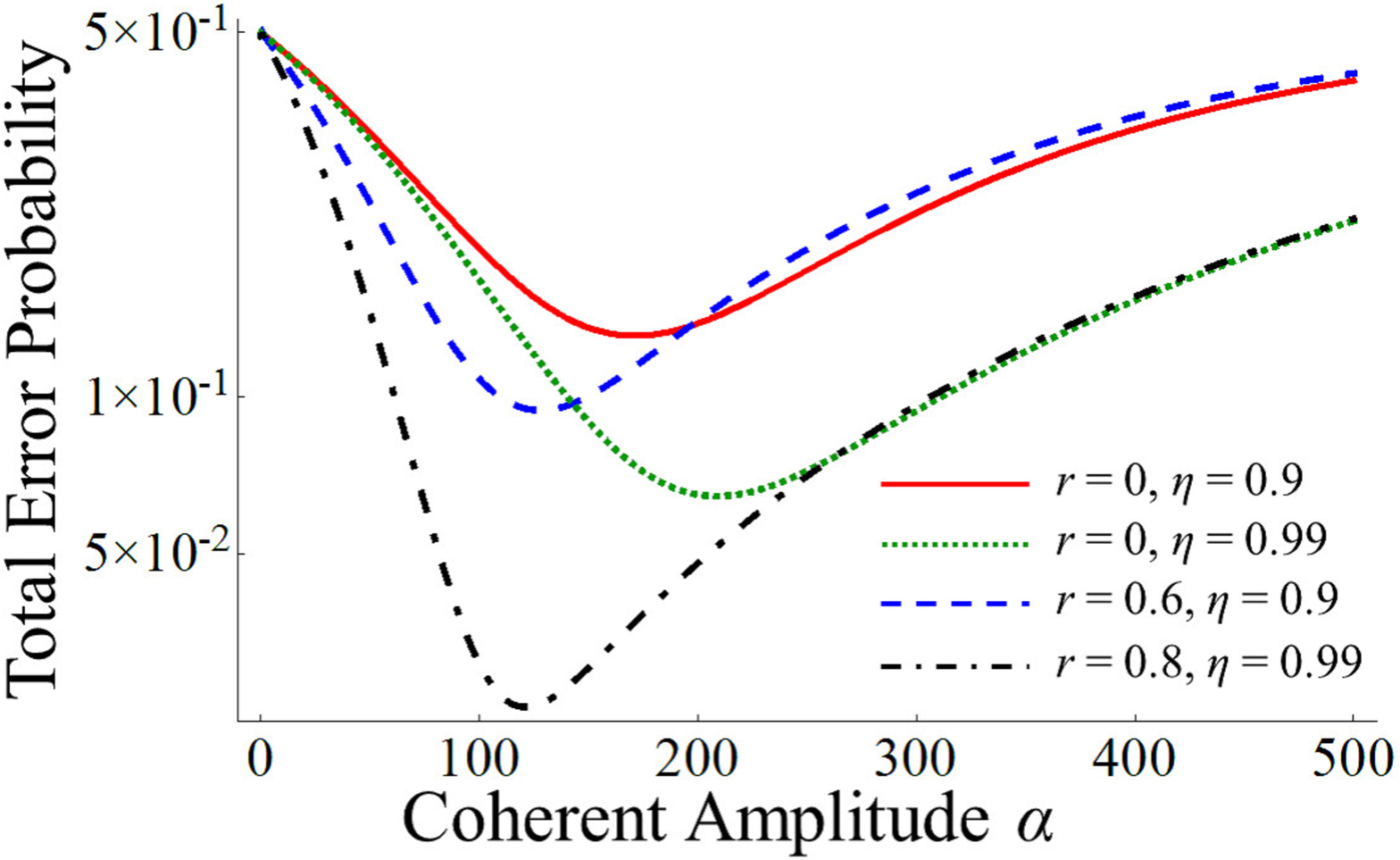}  
\caption{\label{fig7} (Color online) 
Optimization of the error probability of entanglement generation as a function of   with the following values: amplitude of the squeezing parameter $r=0$ and transmittance $\eta=0.9$ (solid line), $r=0$ and $\eta=0.99$ (dashed line), $r=0.6$ and $\eta=0.9$ (dotted line), and $r=0.8$ and $\eta=0.99$ (dot-dashed line).}
\end{figure}
\begin{figure} [htbp]
\includegraphics[width=7cm]{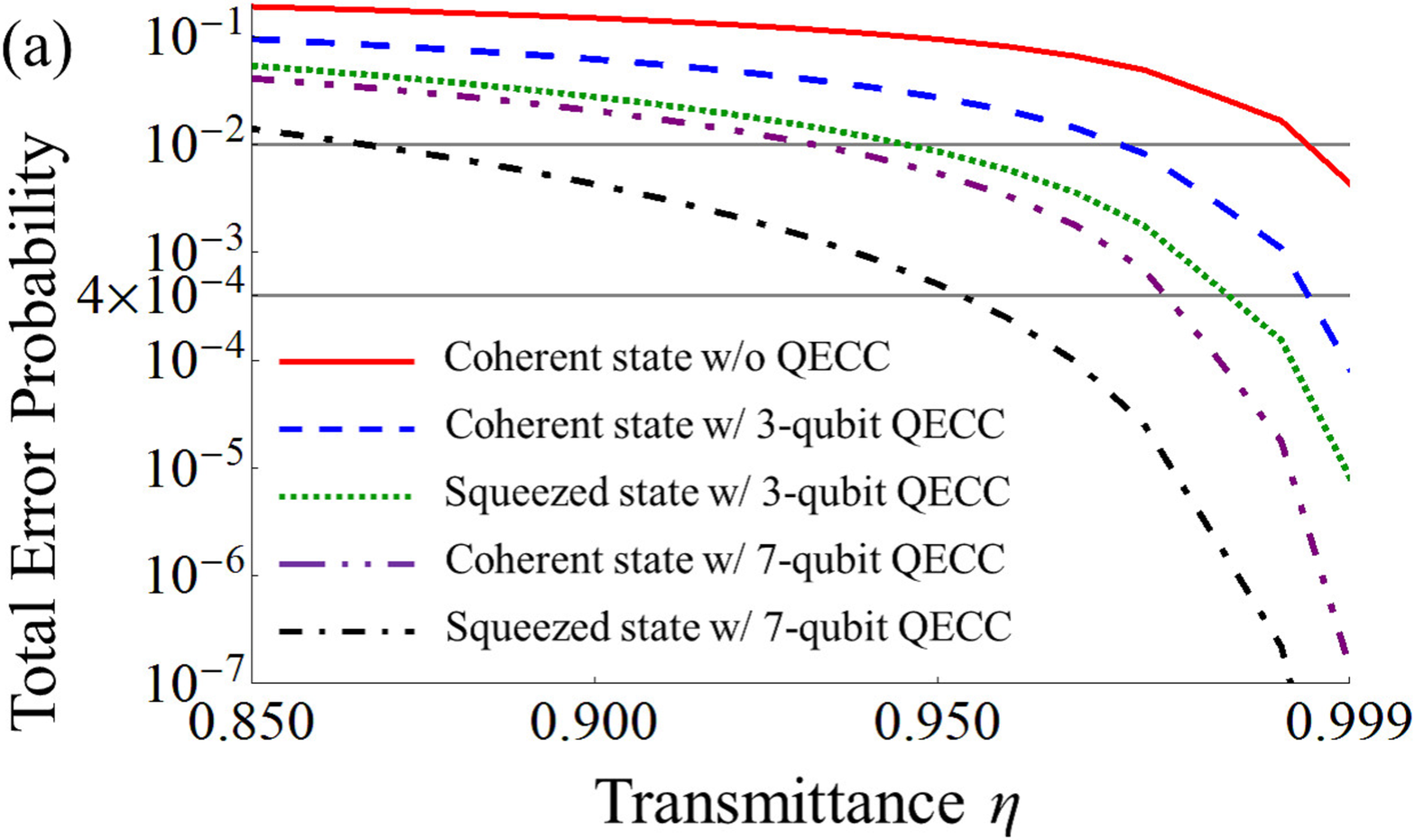}  \\
\includegraphics[width=7cm]{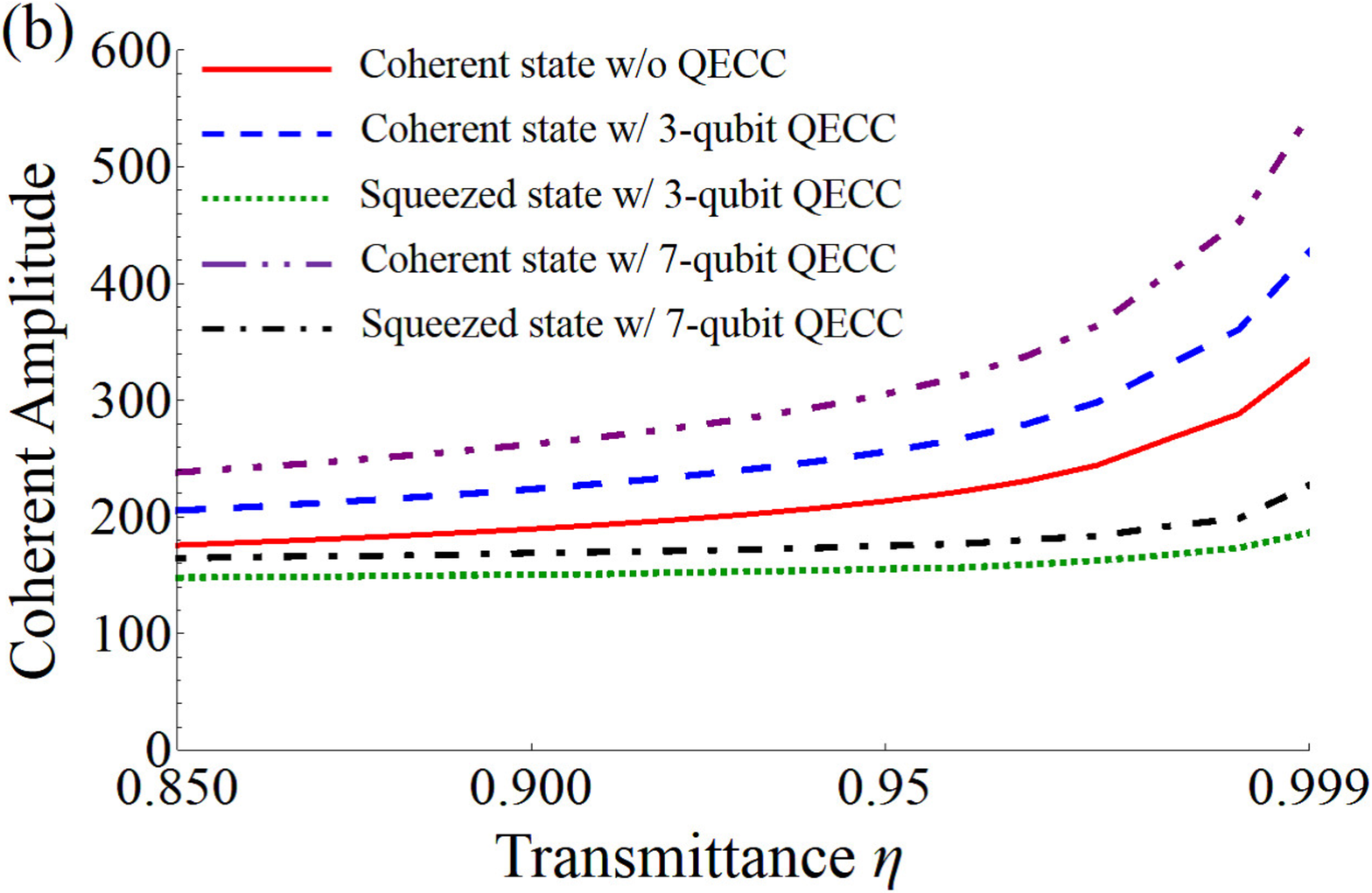}  \\
\includegraphics[width=7cm]{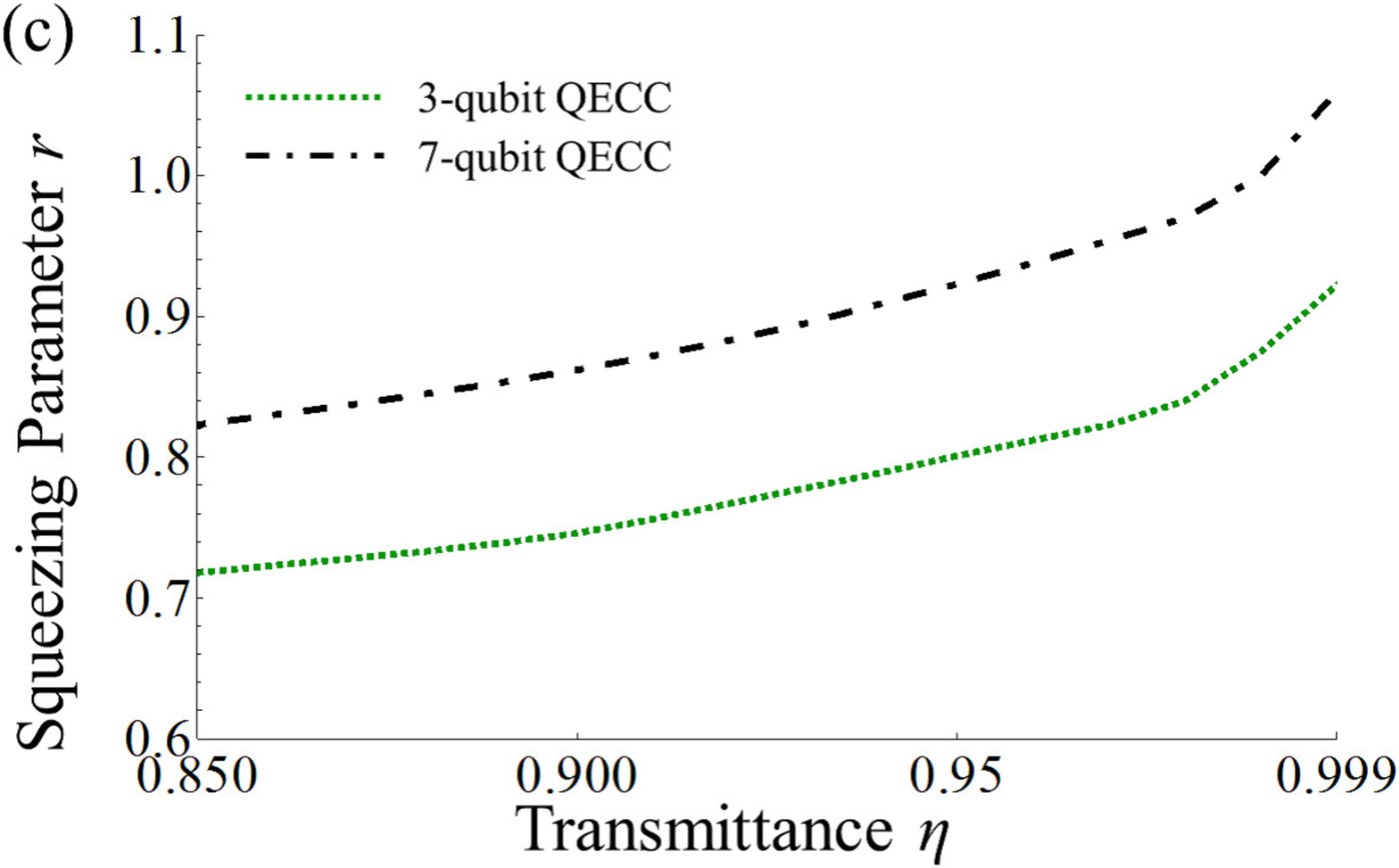}
\caption{\label{fig8} (Color online) (a) The optimized total error probability of entanglement generation for coherent states without QECC (solid line), coherent states with a three-qubit  QECC (dashed line), phase-squeezed states with a three-qubit QECC (dotted line), coherent states with a seven-qubit QECC (dot-dot-dashed line), and phase-squeezed states with a seven-qubit QECC (dot-dashed line). Optimal conditions for the coherent amplitude and squeezing parameter are plotted in (b) and (c), respectively.}
\end{figure}
Then, the total error probability can be obtained by substituting Eqs. (13) and (14) or (15) and (16) into Eq. (12). Figure 8(a) plots the optimized (minimized) error probability for $\theta=0.01$ as a function of the transmittance $\eta$ for coherent states without QECC (solid line), coherent states with a three-qubit QECC (dashed line), phase-squeezed states with a three-qubit QECC (dotted line), coherent states with a seven-qubit QECC (dot-dot-dashed line), and phase-squeezed states with a seven-qubit QECC (dot-dashed line), while (b) and (c) plot the optimized values of the coherent amplitude and squeezing parameter as a function of $\eta$, respectively. Note that to make a fair comparison of phase-squeezed light with coherent light,  since the mean photon number is $\bar{n}=\left| \alpha \right| ^{2}+\sinh{r}$ \cite{31}, we set the coherent amplitude of the phase-squeezed light to $\beta=\sqrt{\left| \alpha \right| ^{2}-\sinh{r}}$. The vertical axis of Fig. 8(b) refers to the coherent amplitudes $\alpha$ and $\beta$ for the coherent states and the squeezed states, respectively. The upper line at $1\times 10^{-2}$ in Fig. 8(a) denotes the threshold of Knillfs method. We also depict the threshold of the FT method by the lower line at $4\times 10^{-4}$ in Fig. 8(a). For the error probability below this threshold, the FT method can also be implemented.\par
In a coherent state without QECC, Knillfs method can be implemented for $\eta\geq 0.999$ (a loss of 0.004 dB). In an optimized phase-squeezed state without QECC, Knillfs method can be implemented for $\eta\geq 0.993$ (a loss of 0.03 dB). In both states without QECC, the FT method is difficult even if $\eta=0.999$. On the other hand, by using a seven-qubit QECC and optimized parameters, Knillfs method and the FT method can be implemented for $\eta\geq 0.94$ (a loss of 0.27 dB) and $\eta\geq 0.99$ (a loss of 0.04 dB) using a coherent state, or for $\eta\geq 0.87$ (a loss of 0.60 dB) and $\eta\geq 0.954$ (a loss of 0.20 dB) using a phase-squeezed state. Since the phase flip error is decreased with more qubits in the QECC, larger coherent amplitudes and squeezing parameters can be used for optimization. Although this QECC for phase flip error is simple, it requires increased atomic resources. Nonetheless, since this increase is polynomial, three- and seven- (or more) qubit QECCs are feasible.

\subsection{Estimation of photon losses} 
Here, we estimate the experimentally feasible values for the total loss. The quantum efficiency of Si photodiodes at 860 nm is 0.93 \cite{40}, which produces a photon loss in the homodyne measurement of 0.07. A smaller coupling loss, compared with conventional Fabry--P$\acute{{\text e}}$rot cavities, is predicted to be less than 0.05 using microtoroidal resonators, which were developed to achieve strong coupling in an atom-cavity system \cite{41,42}. 
Although the predicted coupling loss has not yet been realized, progress in fabrication technology will likely achieve this in the near future. Therefore, the microtoroidal resonator is also a strong candidate for the atom-cavity system with dispersive interaction. A combination of Si photodiodes and microtoroidal resonators provides a total loss as small as 0.1-0.2 (0.46-0.97 dB). This implies that Knillfs method can be implemented using phase-squeezed states with the seven-qubit QECC.

\section{CONCLUSION}
In summary, we have calculated the error probability of entanglement generation by phase-squeezed light to implement fault-tolerant quantum computing methods. When coherent light is used for the qubus, the implementation of any fault-tolerant quantum computation scheme is difficult for practical photon losses, even if a seven-qubit QECC is used, since the error probability exceeds 0.01. In contrast, when phase-squeezed light is used for the qubus with a seven-qubit QECC, Knillfs method can be implemented. With respect to photon losses, when an optimized (coherent amplitude and squeezing parameter) phase-squeezed state is used, Knillfs method can be implemented with up to a 0.24-dB loss (with a three-qubit QECC) or up to a 0.60-dB loss (with a seven-qubit QECC). This result suggests that fault-tolerant quantum computing could be performed using near-future technology, such as the implementation of a seven-qubit QECC. Note that Knillfs method requires repeated operations since the method requires deterministic (or nearly deterministic) two-qubit gates.\par
On the other hand, the FT method can be used with probabilistic two-qubit gates such as the qubus scheme so that repeated operations are not necessary. However, since a 0.03-dB loss (with a three-qubit QECC) or a 0.20-dB loss (with a seven-qubit QECC) are required for the implementation of the FT method, the suppression of photon losses and the implementation of a QECC with more qubits are required.\par
To conclude, we list issues for future study. First, we mentioned in Sec. II that the photon loss does not in principle affect the phase shift operation. However, in practical systems such as the qubus scheme, the decrease of the photon number and squeezing amplitude may affect the phase shift angle at the second cavity. Therefore, we should calculate the error probability, including the difference of photon numbers, or including the displacement operation before the second cavity. Second, although we calculated the error probability for the homodyne measurement, the minimum error discrimination provides the optimal error probability for this qubus scheme, as shown in our previous paper \cite{16}. Therefore, the minimum error probability with photon loss could be derived. Third, the realization of the minimum error measurement for three phase-squeezed states is itself a challenging research task. Fourth, we assumed a phase-squeezed light with large coherent amplitudes to show that the qubus entangler using phase-squeezed light works with a low error probability. In principle, such phase-squeezed light can be obtained by displacing the vacuum squeezed state. However, displacement from a squeezed vacuum to a largely phase-squeezed light pulse with a coherent amplitude of $\beta\gg 1$ has not yet been reported experimentally. The realization of a phase-squeezed light pulse with large coherent and squeezing amplitudes is thus a future research task \cite{43}.

\bibliography{up3bib}

\end{document}